\documentclass[manuscript]{aastex}

\shorttitle{Sunquake in C-class Flare}
\shortauthors{Sharykin and Kosovichev}

\begin{document}

\title{Energy Release and Initiation of Sunquake in C-class Flare}

\author{I.N. Sharykin\altaffilmark{1,2}, A.G. Kosovichev\altaffilmark{1,3,4} and I.V. Zimovets\altaffilmark{2}}
\affil{Big Bear Solar Observatory, New Jersey Institute of Technology,
    Big Bear City, CA 92314, U.S.A}

\altaffiltext{1}{Big Bear Solar Observatory}
\altaffiltext{2}{Space Research Institute (IKI) of the Russian Academy of Science}
\altaffiltext{3}{Stanford University}
\altaffiltext{4}{NASA Ames Research Center}


\begin{abstract}
We present analysis of C7.0 solar flare of Febrary 17, 2013, revealing a strong helioseismic response (sunquake) caused by a very compact impact in the photosphere. This is the weakest known C-class flare generating a sunquake event. To investigate possible mechanisms of this event, and to understand the role of accelerated charged particles and photospheric electric currents, we use data from three space observatories: Ramaty High Energy Solar Spectroscopic Imager (RHESSI), Solar Dynamics Observatory (SDO) and Geostationary Operational Environmental Satellite (GOES). We find that the photospheric flare impact does not spatially correspond to the strongest HXR emission source, but both of these events are parts of the same energy release. Our analysis reveals a close association of the flare energy release with a rapid increase of the electric currents, and suggests that the sunquake initiation is unlikely to be explained by the impact of high-energy electrons but may be associated with a rapid current dissipation or a localized impulsive Lorentz force.
\end{abstract}

\keywords{Sun: flares, chromosphere, magnetic fields, helioseismology, X-rays}

\section{Introduction}

One of interesting effects produced by flare energy release in the solar atmosphere is excitation of helioseismic waves, so-called sunquakes \citep{Kosovichev1998}. Such waves usually propagate as expanding ripples from local impact sources occupying several pixels in photospheric Dopplergrams. The cause of these events is a subject of intensive debates \citep[e.g.][]{Donea2011,Kosovichev2014}. Generally, the necessary condition for producing sunquakes is a sudden momentum enhancement in the lower solar atmosphere. One of the possible agents of such disturbance can be chromospheric heating due to injection of accelerated charged particles postulated by the standard model of solar flares \citep[for a recent review see][]{Fletcher2011}. Models of the gas dynamics processes induced by nonthermal electron beams \citep[e.g.][]{Kostiuk1975, Fisher1985, Kosovichev1986} predict formation of a shock wave or a chromospheric condensation moving towards the solar photosphere and, thus, transferring momentum to the dense plasma. \cite{Kosovichev1995} discussed such beam-driven mechanism of sunquakes. However, the plasma momentum transfer is also possible by other mechanisms, such as sharp enhancement of pressure gradient due to flux-rope eruption \citep[e.g.][]{Zharkov2013} or by an impulse of the Lorenz force which can be stimulated by electric currents in the lower solar atmosphere \citep{Fisher2012}. Also, it is possible that different sunquake events are caused by different mechanisms. Usually, sunquakes are associated with M and X class flares. However, many X-class flares did not produce sunquakes \cite[e.g.][]{Donea2011}, whereas these events had been noticed during relatively weak M-class flares \citep{Martinez-Oliveros2008, Kosovichev2014}.

In this Letter, we discuss observations of the C7.0 flare of February 17, 2013, which produced a rather strong sunquake initiated during the HXR burst. We use data from four space instruments: EUV observations from SDO/AIA \citep{Lemen2012}, vector magnetic field measurements from SDO/HMI \citep{Scherrer2012}, integrated soft X-ray emission from GOES, and X-ray spectroscopic imaging data from RHESSI \citep{Lin2002}. We investigate potential mechanisms of the sunquake initiation, and find that despite the precise temporal coincidence between the HXR impulse and the photospheric impact this event is not consistent with the standard flare model, because the HXR source and the sunquake impact were at different spatial locations, at two different footpoints of the flare loop. Our analysis leads to a suggestion that a significant role in the sunquake initiation may be played by electric currents in the low atmosphere.

\section{General description of the event and sunquake}

The flare event of Febrary 17, 2013, was observed in active region NOAA 11675. It consists of two  subflares clearly separated in time and space: the first subflare has the C7.0 GOES X-ray class, and the second subflare reached M1.9 peak intensity. The duration of the whole double flare is about 8 min, starting at 15:46:00 UT and ending at 15:54:00 UT (Fig. 1). The highest energy of the hard X-ray (HXR) emission (maximum at 15:47:20 UT), detected by RHESSI during the first subflare is $\sim 1$ MeV. The second subflare is characterized by weaker intensity and energy ($<$300 keV) of HXR emission which reached maximum at 15:50:30 UT.

Top panels of Fig. 2 present the AIA images in the 94 $\rm\AA$ channel. The temporal and spatial resolutions are  12 seconds and $1.2^{\prime\prime}$ (with the angular pixel size of $0.6^{\prime\prime}$). The preflare state reveals a compact loop-like structure where the flare process occurs.

The sunquake is observed as an expanding circular wave in the HMI Dopplergrams filtered in the frequency high range, 5-6 mHz, to isolate the sunquake signal from the convective noise. The propagation of the sunquake wave is shown on the time-distance diagram \citep{Kosovichev1998,Zharkova2007} presented in Fig.2 (middle right panel) comparing the observed signal with the theoretical ray-path theory prediction (dotted yellow line). The time-distance analysis shows that the sunquake is initiated by the first subflare with the initiation point at the flare impulse signal. The inclined wave pattern above the theoretical curve is associated with the frequency dispersion of the helioseismic wave packet.

The initiation of the sunquake is observed as a strong localized impulse in the HMI Dopplergrams and line-of-sight magnetogram at $\approx$15:47:54 UT. The location of the impulse on the HMI magnetogram is illustrated in Fig. 2 (middle left panel). Because of the rapid variations during the flare impulse the Doppler velocity and magnetic field measurements in the impact pixels can be inaccurate. Therefore, we use the original level-1 HMI filtergram data from the two HMI cameras to locate the exact time and place of the flare impact. The bottom panels of Fig. 2 shows the time difference between the HMI filtergram images (from HMI Camera-1). We see that compared to the preflare time during the flare there is enhancement of emission in the pixels associated with the AIA brightnings and the place of the sunquake initiation. The timing of the photospheric impact is illustrated in Fig. 1 (bottom) which shows the signals from both HMI cameras as a function of time. The periodic variations of these curves are due to the line scanning. The plot shows that the photospheric impact coincides with the HXR impulse within 3 sec (the HMI camera resolution).

\section{Spatial structure of the flare region}

Here we present a description of the spatial structure of the flare region according to the RHESSI and AIA/SDO observations. RHESSI uses a Fourier technique to reconstruct X-ray emission sources \citep{Hurford2002}. We apply the CLEAN algorithm to synthesize the X-ray images using detectors 1,3-6,8 (integration times are shown in Fig.3). In Fig. 3, the RHESSI HXR and SXR contour images are compared with the corresponding AIA 94 \AA ~images for the time interval covering the HXR peaks of both flares. To compare positions of the EUV and X-ray sources with the structure of magnetic field we plot the polarity inversion line from the HMI magnetogram. The structure of the EUV emission sources is rather complicated. There are ribbon-like structures
located on both sides of the magnetic field inversion line. During the HXR burst we observe a loop-like structure with one footpoint associated with very strong HXR emission (25-200 keV), and the other footpoint located in the place of the photospheric impact (sunquake initiation), which also coincides with a weak HXR emission source. The total emission intensity of the weaker X-ray source is approximately five times less than the emission intensity of the stronger HXR source. If the sunquake were initiated by an impact of high-energy electrons, then their impact would be in the place of the intensive energy loss of the accelerated particles, and coincide with the strongest HXR emission source. However, we observe the opposite situation when the sunquake impact correlates with a weaker HXR emission source. This indicates that the sunquake is unlikely be generated by the impact of high-energy electrons.

The second subflare has SXR source (6-12 keV) coinciding with the HXR source (25-50 keV) and saturated UV emission above the magnetic field inversion line. However, this subflare is located $\sim$3 Mm away from the place of energy release in the first subflare and according to our analysis is not associated with the sunquake.

\section{Analysis of RHESSI spectra: accelerated particles and heating}

To determine properties of the accelerated particles, the plasma and their energetics we use the RHESSI data in the range 5-250 keV. We investigate two spectra taken during the HXR peaks of two subflares. The power-law approximation $f(E)=AE^{-\gamma}$ ($A$ is normalization coefficient) is considered for the hard X-ray (HXR) nonthermal emission $\gtrsim 20$ keV. To simulate the presence of the low-energy cutoff we use the broken power law \citep{Holman2003} with fixed photon spectral index $\gamma_0=1.5$  below the break energy ($E_{low}$). For the first subflare, an additional break energy ($E_{br}$) is considered, and, thus, for this case we have two spectral indices $\gamma_1$($E_{low}<E<E_{br}$) and $\gamma_2$($E>E_{br}$). For the second subflare we consider only one spectral index $\gamma_1$($E>E_{low}$) and also make a pileup correction as the count rate is sufficient to observe such effect.

The thermal soft X-ray (SXR) spectrum $\lesssim 20$ keV is approximated by one-temperature thermal bremsstrahlung emission with two parameters: temperature ($T$) and emission measure ($EM$). The RHESSI spectra are fitted by means of the least squares technique implemented in the OSPEX package with 7 free parameters ($EM$, $T$, $A$, $E_{low}$, $\gamma_1$, $E_{br}$ and $\gamma_2$) for the first subflare and 5 parameters ($EM$, $T$, $A$, $E_{low}$ and $\gamma_1$) for the second subflare. Fig. 4 displays results of the fitting.

From the thermodynamics point of view the second subflare is hotter than the first one, but the emission measure is smaller. Volume $V$ of the UV loop estimated in the previous section is $10^{26}$ cm$^3$, so that plasma density $n_1 = \sqrt{(EM_1/V)}\approx 6\times 10^{10}$ cm$^{-3}$ for the first subflare. The plasma density for the second subflare, assuming the same flare region volume, is $n_2\approx 2\times 10^{11}$ cm$^{-3}$. Due to compactness of the flare region the plasma within the magnetic loops is rather dense.

The HXR photon spectrum is harder for the first HXR burst than for the second one. Normalization coefficient $A$ of the HXR spectrum is also one order of magnitude larger in the case of the first subflare. This means that the acceleration process is more efficient during the first subflare. The total flux $Fl$ [electrons s$^{-1}$] of accelerated electrons can be estimated following the work of \cite{Syrovatskii1972}:

$$
Fl(E_{low}<E<E_{high}) = 1.02\times 10^{34}\frac{\delta_1^2}{E_{low}\beta(\delta_1,1/2)}\frac{I_{ph}(E_{low}<E<E_{high})}{[1-(E_{low}/E_{high})^{\delta_1}]}
$$

where $E_{high}$ is the upper energy cutoff, $\delta_1 =\gamma_1 - 1 $ is the spectral index of accelerated electrons in the HXR emission region, $\beta(x,y)$ is beta function, and $I_{ph}(E_{low}<E<E_{high})$ photons s$^{-1}$ cm$^{-2}$ is the energy integrated photon spectrum in the range shown in the brackets. From the fitting results using this formula, we obtain electron fluxes $Fl_1\approx (2.0\pm 1.2)\times 10^{35}$ and $Fl_2\approx (0.10\pm 0.06)\times 10^{35} $ electrons s$^{-1}$ for the first and second subflares. Despite the lower GOES class of the first subflare we observe more energetic electrons involved in the acceleration process of this subflare than in the second one. Theoretically, these electrons could contribute to the sunquake initiation. However, the discrepancy between the locations of the sunquake impact and the strongest HXR emission source indicates that the beam-driven origin of the sunquake is unlikely.

\section{Electric currents in the flare region}
Local heating by electric currents or impulsive Lorentz force can also be a source of the sunquake. In this section we consider the evolution of electric currents at the photosphere level. To estimate the horizontal electric currents we use the Faraday law applied to the 45-second line-of-sight HMI magnetograms with the spatial resolution $1^{\prime\prime}$ and pixel size $0.5^{\prime\prime}$:

$$
\oint_C\vec{E}\cdot\vec{dl} = -\frac{1}{c}\frac{d}{dt}\left(\int_{S_C}\vec{B}\cdot \vec{dS}\right)
$$

We can estimate the average transversal component of the electric field $<E_{\perp}>=[dF_z/dt]/cL$, where $F_z$ is total magnetic flux inside a contour with length $L$, which covers the flare region. The evolution of $<E_{\perp}>$ presented in Fig.1 (gray histograms in top panels) shows that both subflares correlate with the peaks of $<E_{\perp}>$.

To calculate the vertical currents we use the disambiguated HMI vector magnetic field data \citep{Centeno2014} with time cadence 720 seconds, and the same spatial resolution as of the line-of-sight magnetograms. The vertical electric current density is calculated from the HMI vector magnetograms using the Ampere's law \citep[e.g.][]{Guo2013}:

$$
j_z=\frac{c}{4\pi}(\nabla\times\vec{B})_z = \frac{c}{4\pi}\left(\frac{\partial B_x}{\partial y} - \frac{\partial B_y}{\partial x}\right)
$$

The resulted $j_z$ map during the flare, effectively averaged over 12 min due to the HMI temporal resolution, is presented in Fig.~3. Figure 5 displays the evolution of $<j_z>$ averaged through the flare region with area $\approx 1.5\times 10^{18}$ cm$^2$, and reveals a maximum corresponding to the flare. We estimated errors for $<j_z>$ as the standard deviation of the $j_z$ distribution in the quite Sun regions.

In Fig.~3 we see that the place of sunquake generation correlates with the strong electric currents, and that there is no significant HXR emission in this place. The HXR is mostly emitted from the source located on the other side of the magnetic field polarity inversion line at the opposite footpoint of the flare loop. Such observation, and the time evolution of $<j_z>$ and $<E_{\perp}>$ can be an evidence of a non beam-driven origin of sunquake. The correlation with the location of the strongest electric currents suggests that the sunquake event could be initiated due to a local heating or impulsive Lorentz force in the flare region.


\section{Discussion}

In this section we will discuss the contributions of the electric currents and nonthermal electrons in flare energy release and the generation of the sunquake.

For the estimated fluxes of accelerated electrons for the first subflare, the total kinetic power $P_{nonth}\approx 1.5\times 10^{27}$ erg s$^{-1}$ in the HXR peak. To estimate the Joule heating in the sunquake generation region we need to estimate effective electric conductivity $\sigma_{eff}$. In the regime of electric currents dissipation the magnetic Reynolds number $Re_m=4\pi\sigma_{eff}L^2/(c^2\tau)\sim 1$, where $\tau$ is a characteristic time of electric current dissipation ($\sim 100$ s, the duration of the HXR burst), and $L$ is a characteristic length scale ($\sim 1^{\prime\prime}$, the size of the impulsive region on the Dopplergrams). For these characteristic values we get $\sigma_{eff}\sim 10^6$ CGS units. This value is substantially higher than the theoretical Spitzer conductivity \citep{Kopecky1966}. However, recent studies of the partially ionized plasma of the solar chromosphere show that the electric conductivity can be substantially reduced due to Pedersen resistivity \citep[e.g.][]{Leake2012} or due to small scale MHD turbulence \citep{Vishniac1999}. The volumetric energy release is $Q_j = j^2/\sigma_{eff}\approx 8\times 10^3$ erg s$^{-1}$ cm$^{-3}$ for $j\approx 0.3$ A/m$^2$. The total energy release due to dissipation of electric currents in the sunquake region is $Q_j^{tot}\approx 3\times10^{27}$ erg s$^{-1}$, estimating the volume for a box with length scale $L\sim 1^{\prime\prime}$. So, we see that $P_{nonth}\sim Q_j^{tot}$, and both types of energy release have the energy budget sufficient to explain heating in the flare according to the GOES data: the change of the plasma internal energy, $d(3nk_BTV)/dt\sim 10^{27}$ erg s$^{-1}$, and the radiation losses, $L_{rad}\sim 5\times 10^{26}$ erg s$^{-1}$.

To produce the sunquake we need a strong impulsive force in the lower solar atmosphere. The sunquake momentum can be estimated from the initial impact as $p_{sq}\sim \rho L^3v\sim 10^{22}$ g$\cdot$cm s$^{-1}$ for $\rho\sim 10^{-8}$ g$\cdot$cm$^{-3}$ (photospheric value) and $v\sim c_s\sim 10$ km/s, where $c_s$ is photospheric sound speed. In principle, the force generating sunquakes can be produced directly by energetic electron beams. The total momentum of injected nonthermal electrons is

$$
p_e=\tau\sqrt{2m_e}\int_{E_{low}}^\infty f(E)\sqrt{E}dE
$$

where $m_e$ is the electron mass, $f(E)$ is the distribution function of nonthermal electrons, and $\tau$ is the characteristic time of the injection. For the first HXR burst, $p_e\sim 10^{20}$ g$\cdot$cm s$^{-1}$. As the emission intensity of the weaker HXR source associated with sunquake impact is five times less comparing with strong HXR source, than nonthermal electrons momentum in the footpoint associated with sunquake impact $\sim 0.2\times 10^{20}$ g$\cdot$cm s$^{-1}$.

Momentum of the accelerated protons can be much larger than in the case of electrons and lead to stronger disturbances in the solar atmosphere. Assuming that the protons roughly (not accounting collisions) have energy $E_p\lesssim E_e$, the momentum contained in the proton beam $p_p\lesssim p_e\sqrt{(m_p/m_e)}\sim 45p_e\sim 0.5\times 10^{22}$ g$\cdot$cm s$^{-1}$. We see that the momentum of accelerated protons represents a more probable agent of the sunquake initiation than the momentum of electrons.

Our observations show that while the sunquake impact and the HXR impulse are simultaneous in time they are clearly separated in space, and located at the different footpoints of the flare magnetic loop. In addition, we find that the impact location correlates with the strongest electric currents. This suggests that, perhaps, energetic particles are accelerated by electric field in the place of sunquake initiation, and then the particles travel along the flare magnetic loop to the other footpoint and caused the HXR emission.

The impulsive plasma motion in the lower solar atmosphere may be caused by fast heating due to Joule dissipation or sharp increasing of the Lorentz force. In the first case we can estimate the plasma momentum as $p_J\sim \tau V\nabla P\sim P\tau L^2$, where $\nabla P$ is the pressure gradient on the length scale, $L$. The pressure can be estimated from the energy equation

$$
\frac{dP}{dt}=\frac{j^2}{\sigma_{eff}}-L_{rad}
$$

where $L_{rad}$ is the radiation heat loss which is the main source of cooling in the lower solar atmosphere. From this equation $P\lesssim j^2\tau/\sigma_{eff}$ and, hence, $p_J \lesssim (j\tau L)^2/\sigma_{eff}\sim 10^{23}$ g$\cdot$cm s$^{-1}$.

The plasma momentum, associated with the Lorentz force, is $p_L\sim jB\tau L^3/c\sim 10^{22}$ g$\cdot$cm s$^{-1}$, where $B\sim 100$ G is the magnetic field in the sunquake source, and $c$ is the speed of light.

From the estimated values of $Q_j$, $p_J$ and $p_L$ one can conclude that the appearance of strong electric currents in the lower solar atmosphere is sufficient to explain the flare energy release and generation of the sunquake. Moreover, these estimations show that the electric current driven disturbances are sufficiently strong, and also that the electric currents are concentrated in the place of the sunquake initiation while the strongest HXR impulse is $\sim 3$ Mm away. Therefore, it is likely that not only high-energy particles play significant role in the flares, as assumed by the standard flare model, but also electric currents in the lower solar atmosphere can be also a significant part of flare energy release. In our recent paper, we discuss the relationship between electric currents and the fine structure of flare ribbons \citep{Sharykin2014}.

\section{Summary and conclusion}

The main results of the work are as the following:

\begin{enumerate}
\item We observed a strong sunquake event in a weak C-class flare.
\item The sunquake is initiated exactly, within 3 sec observational accuracy, during the burst of the HXR emission.
\item The place of the photospheric impact associated with the sunquake generation corresponds to the weaker HXR emission source, while there is no significant photospheric impact in the stronger HXR emission source, which is located at the opposite footpoint of a flare loop observed in the EUV AIA images.
\item The place of the photospheric impact associated with sunquake initiation corresponds to the most intense electric currents.
\item The total (C7.0-M1.9) flare event temporarily correlates with the maxima of vertical and transversal electric currents estimated in the energy release site.
\end{enumerate}

The main conclusion of the presented observational results is that the helioseismic response (sunquake) and flare energy release in the lower solar atmosphere may have strong connection to photospheric electric currents. The sunquake impact may be initiated by a pressure gradient caused due a rapid current dissipation or impulsive Lorentz force. The discovery of the strong photospheric impact produced by a weak C7 flare, which initiated the helioseismic response, opens new perspectives for studying the flare energy release and transport because such flares usually have relatively simple magnetic topology and do not saturate detectors of space and ground-based telescopes. However, our results show that high spatial and temporal resolutions are needed for these studies.

The work was partially supported by RFBR grant 13-02-91165, President's grant MK-3931.2013.2, NASA grant NNX14AB70G, and NJIT grant.


\clearpage

\begin{figure}
\epsscale{.80}
\plotone{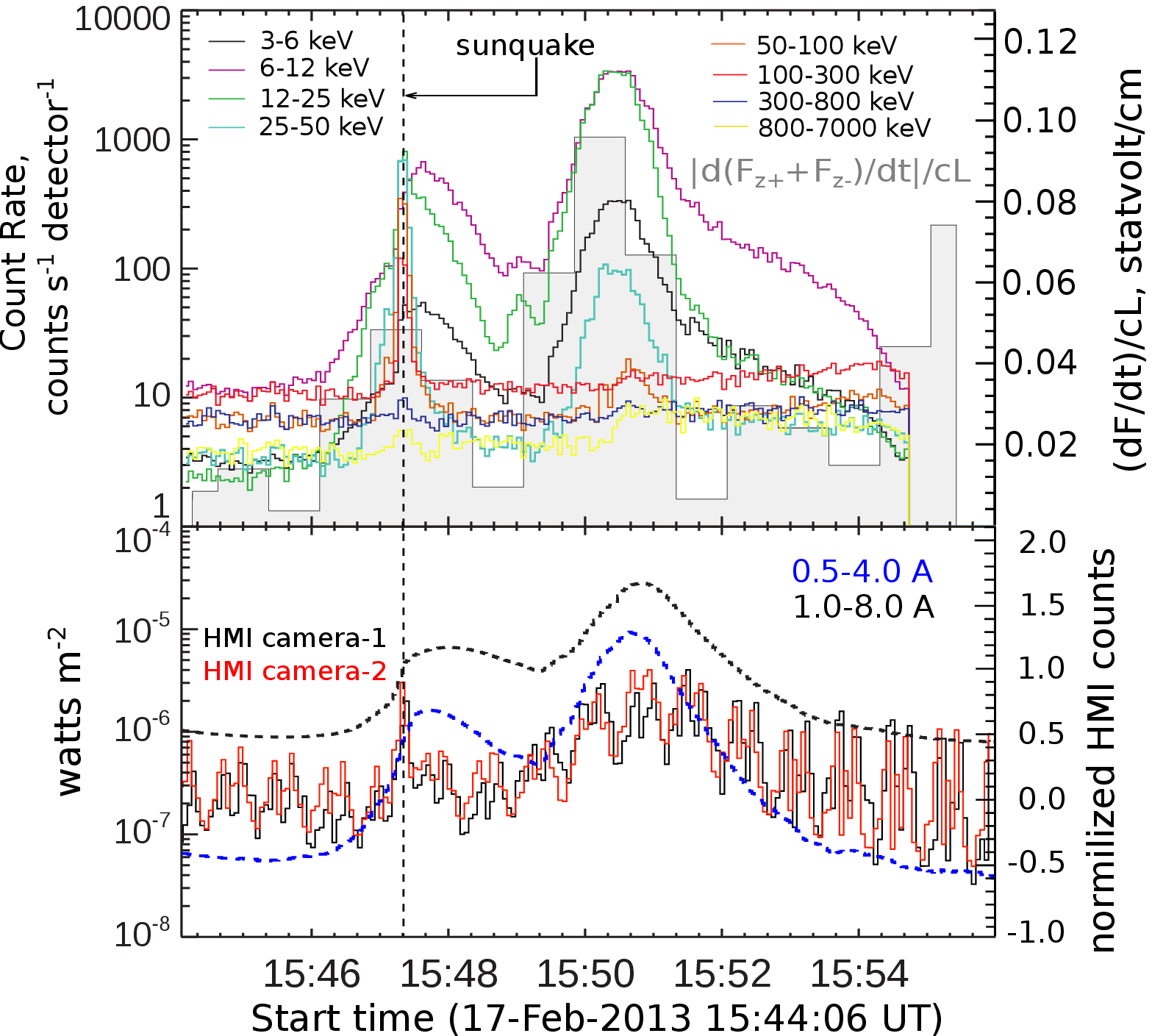}
\caption{Top: RHESSI count-rate in different energy bands (marked in the figure) and effective transversal electric field in the flare region (gray histogram). Bottom: GOES light curves in two channels 0.5-4 $\rm\AA$ (blue dashed line) and 1-8 $\rm\AA$ (black dashed line), and normalized counts from the HMI Camera 1 (black) and Camera 2 (red), revealing the sunquake initiation impulse. \label{fig1}}
\end{figure}

\clearpage
\begin{figure}
\epsscale{0.8}
\plotone{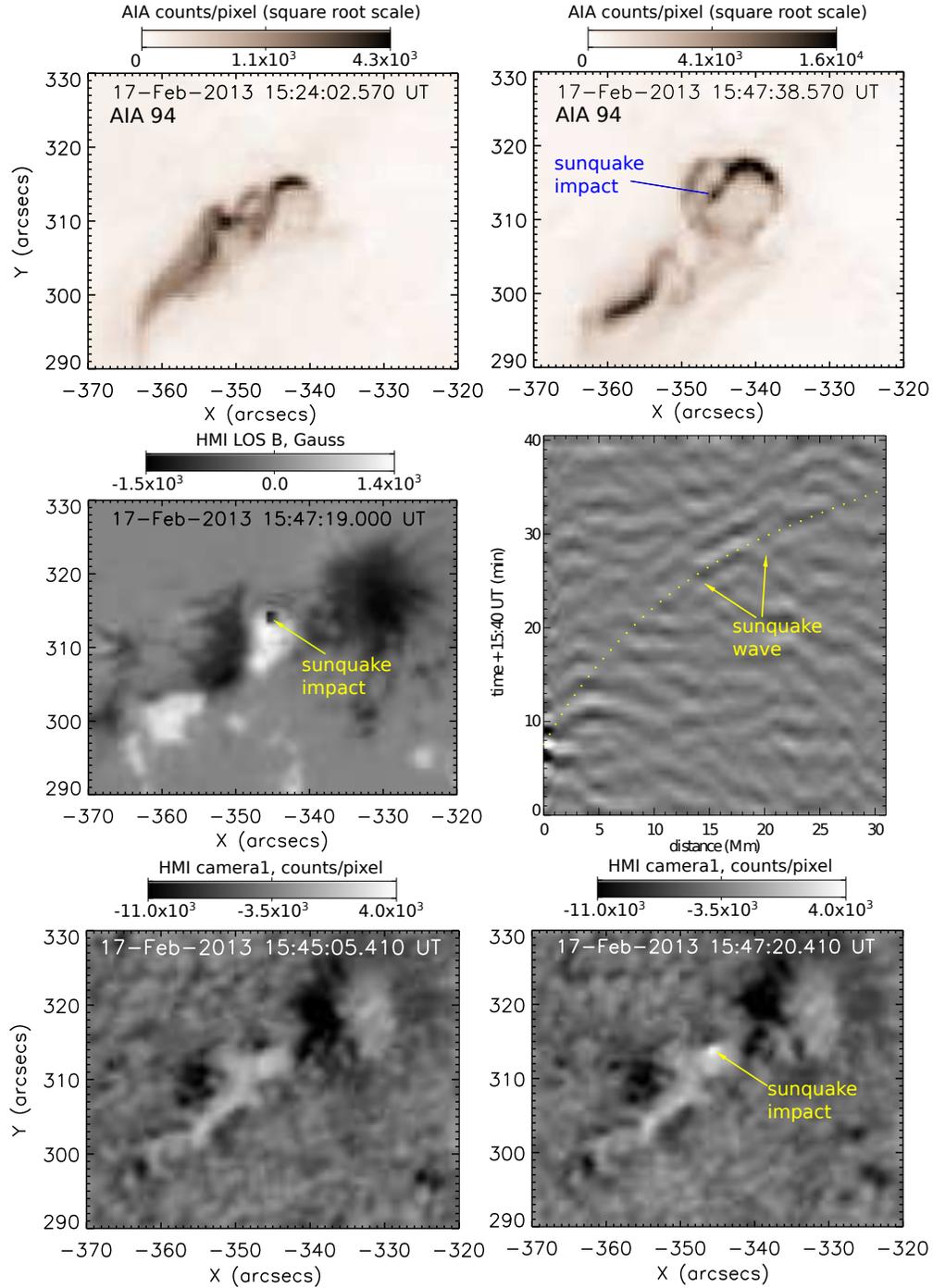}
\caption{Top: AIA 94 \AA~ images before and during the flare. Middle left: HMI line-of-sight magnetogram showing the impact location. Middle right: the sunquake time-distance diagram with the ray-path theoretical prediction (dotted yellow line). Bottom panel: time differences of the HMI level-1 filtergram.\label{fig2}}
\end{figure}

\begin{figure}
\epsscale{1.0}
\plotone{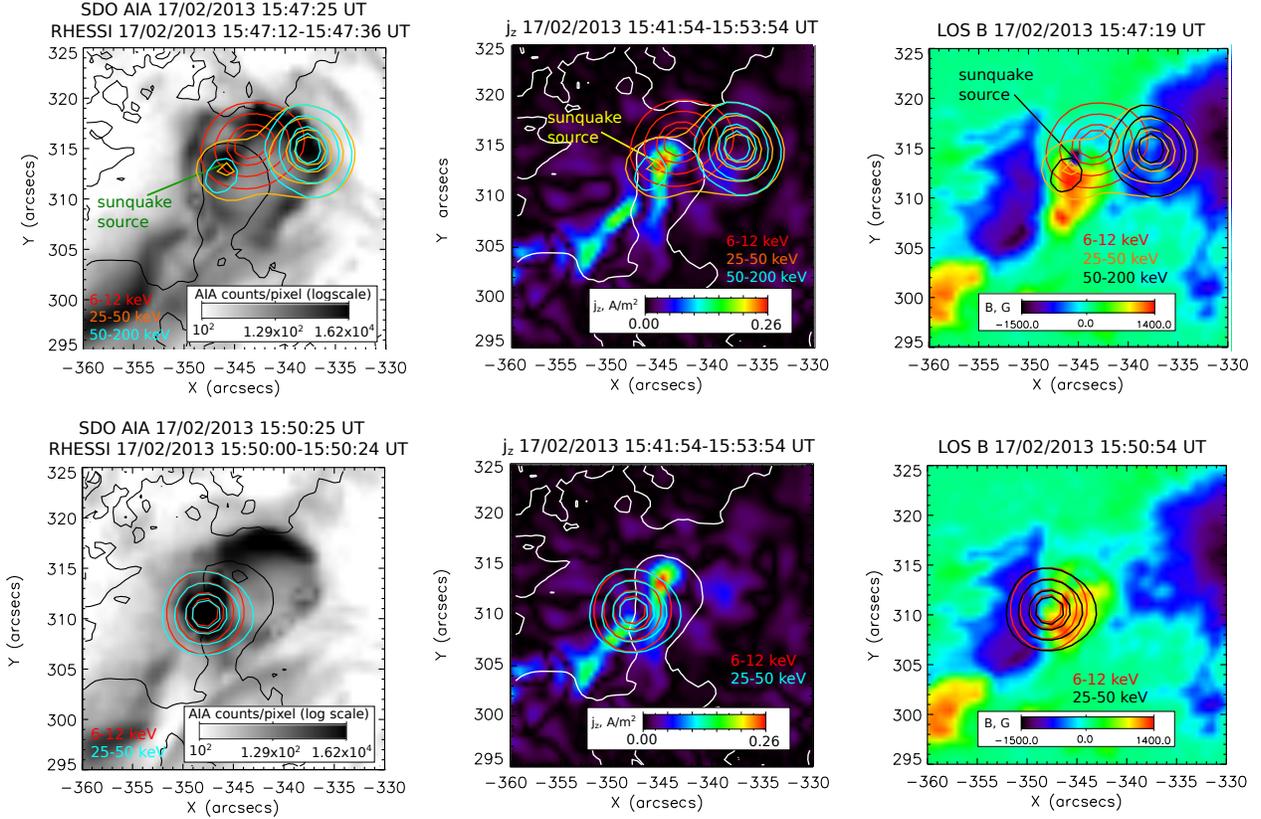}
\caption{Left panels show the EUV AIA 94 $\rm\AA$ images during the two subflares, and the magnetic polarity inversion line (black curves). Middle panels show the magnitude of the vertical electric current density, $j_z$ (obtained from the HMI vector magnetic field measurements for the time interval 15:41:54-15:53:54 UT), and the polarity inversion line (white curves). Right panels show the HMI line-of-sight magnetograms. Color contours (40, 60, 80 and 90 \%
) on the all images show the RHESSI CLEAN-reconstructed maps of the X-ray emission in the energy ranges marked on the plots.\label{fig3}}
\end{figure}

\clearpage
\clearpage
\begin{figure}
\epsscale{1.0}
\plotone{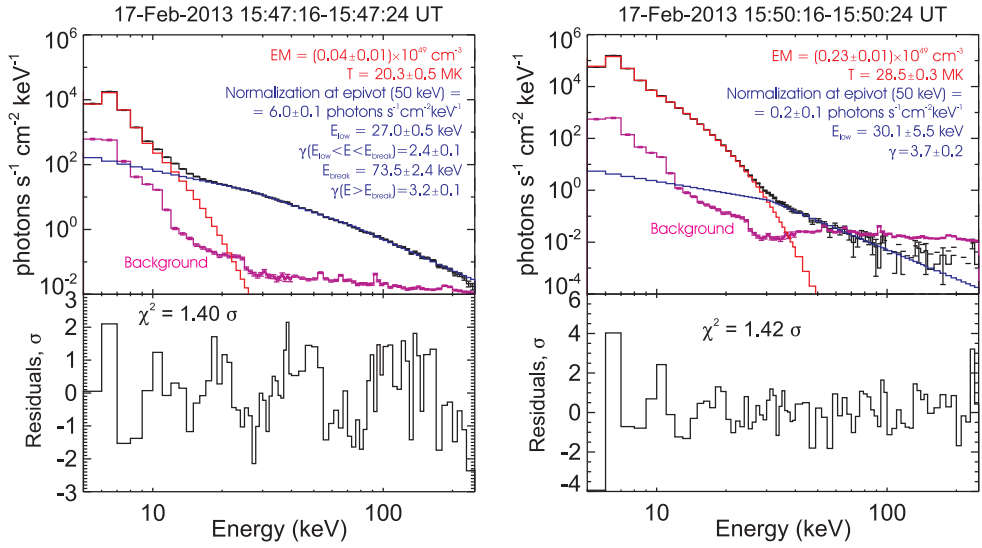}
\caption{ Black line shows the RHESSI X-ray spectra; red line shows the fit of the thermal SXR spectrum using a one-temperature bremsstrahlung emission model; blue line shows the fit of the HXR non-thermal spectrum by a power law function; violet line shows the RHESSI background spectrum. Tables of the fitted parameters are shown in the top panels. \label{fig4}}
\end{figure}

\clearpage
\clearpage
\begin{figure}
\epsscale{1.0}
\plotone{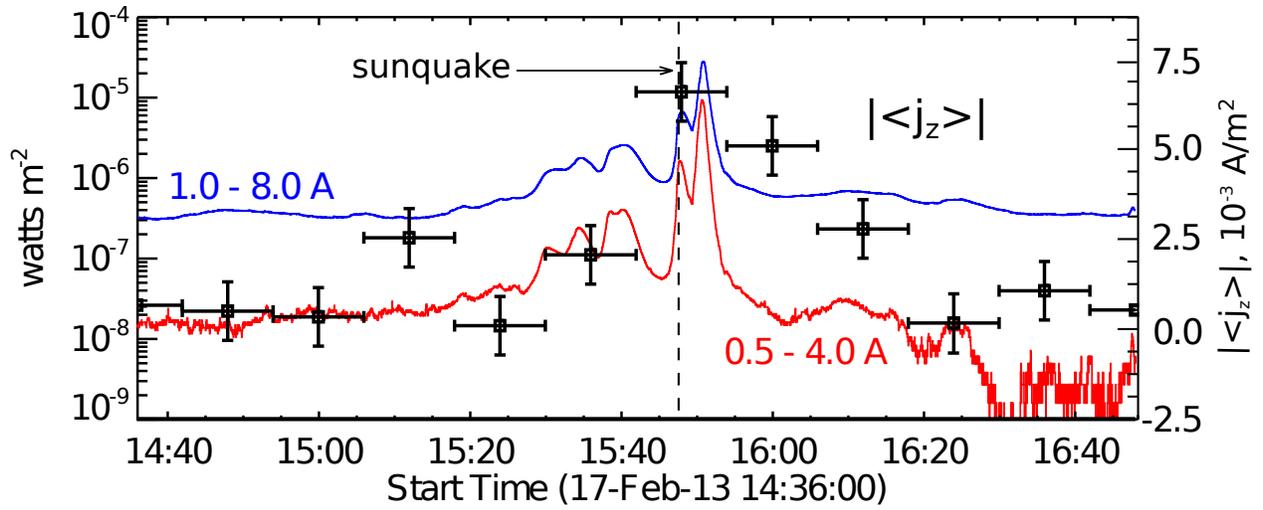}
\caption{GOES light curves in two channels 0.5-4 $\rm\AA$ (red line) and 1-8 $\rm\AA$ (blue line) compared with the evolution of the vertical electric current (squares with error bars) averaged over the flare region. Horizontal bars indicate the temporal resolution of the HMI vector magnetogram measurements. \label{fig5}}
\end{figure}

\clearpage

\clearpage
\end{document}